\documentclass[prl, twocolumn, letterpaper, nofootinbib]{revtex4-2}

\usepackage{amsmath}
\usepackage{amssymb}
\usepackage{graphicx}
\usepackage{hyperref}
\usepackage{color}

\begin{document}

\title{Cosmological perturbations: non-cold relics without the Boltzmann hierarchy}

\author{Lingyuan Ji}
\email{lingyuan.ji@jhu.edu}
\author{Marc Kamionkowski}
\email{kamion@jhu.edu}
\author{Jose Luis Bernal}
\email{jbernal2@jhu.edu}
\affiliation{William H.\ Miller III Department of Physics and Astronomy, Johns Hopkins University\\ 3400 North Charles Street, Baltimore, MD 21218, USA}

\begin{abstract}
We present a formulation of cosmological perturbation theory where the Boltzmann hierarchies that evolve the  neutrino phase-space distributions  are replaced by integrals that can be evaluated easily with fast Fourier transforms.  The simultaneous evaluation of these integrals combined with the differential equations for the rest of the system (dark matter, photons, baryons) are then solved with an iterative scheme that converges quickly.  The formulation is particularly powerful for massive neutrinos, where the effective phase space is three-dimensional rather than two-dimensional, and even moreso for three different neutrino mass eigenstates. Therefore, it has the potential to significantly speed up the computation times of cosmological-perturbation calculations. This approach should also be applicable to models with other non-cold collisionless relics.
\end{abstract}
\maketitle

\paragraph{Introduction} The publicly available
cosmological-perturbation codes {\tt CAMB} \cite{Lewis:1999bs}
and {\tt CLASS} \cite{Lesgourgues:2011rh} lie at the heart of almost
all analyses in cosmology.  These codes solve the differential
equations for the evolution of the gravitational potentials, the
baryon and dark-matter fluid equations, the neutrino and
photon distribution functions, and possibly more species, depending on the cosmological model considered.  The codes, which build upon
nearly half a century of technical innovations \cite{earlywork},
are now remarkably efficient.  However, modern Markov Chain
Monte Carlo (MCMC) analyses require these codes to be called
tens of thousands times to obtain the posterior in a multidimensional cosmological-parameter space, requiring perhaps
days of CPU time.  There is thus incentive to accelerate these codes.

The most time-consuming parts in these calculations are the
``Boltzmann hierarchies'', which evolve the higher moments of
the photon and neutrino distribution functions.  The real
bottleneck, though, are massive neutrinos: since their momentum
distribution occupies a three-dimensional, rather than
two-dimensional, space, they require, strictly speaking, an
\emph{infinitude} of hierarchies.  Nonzero neutrino masses are,
moreover, becoming increasingly important given that they will
be probed with forthcoming cosmological measurements
\cite{Green:2019glg}.  Clever numerical methods are 
able to reduce the system of ordinary differential equations (ODEs) to a manageable size
\cite{Lesgourgues:2011rh}. But the algorithms are still
ultimately limited by requirement to solve---depending on the
target accurac ---$O(500)$ ODEs (for {\it each} Fourier
wavenumber $k$) for the Boltzmann hierarchies of photons and
three generations of massive neutrinos.  The computational
problem is exacerbated further with the increased focus on new-physics models with
other non-cold relics or neutrino models with non-thermal phase-space distributions; we list in
Refs.~\cite{Deramo:2020gpr,Das:2021pof,Kunze:2021qxt,Decant:2021mhj,Abellan:2020pmw,Alvey:2021sji}
papers just the past year on such relics.

It has long been known that each Boltzmann hierarchy is formally
equivalent to a small set of integral equations
\cite{integralequations}, but only recently \cite{Kamionkowski:2021njk} has this formalism
been implemented for scalar perturbations numerically.
Numerical experiments in which the photon hierarchies were
replaced with the integral equations showed that the new
``hierarchy-less'' formalism may have the potential to
accelerate cosmological-perturbation codes.  We emphasize that this formalism provides a numerical solution to the perturbation equations; it is not an analytical approximation.

Here, we apply this integral-equation approach to neutrinos (and
other collisionless non-cold relics) and show that it is
potentially extremely powerful.  First of all, the integral
equations for collisionless particles are simply integrals.
Moreover, each integral can be written as a convolution of
gravitational potentials and a radial eigenfunction, and the convolution can be done trivially with a fast Fourier transform (FFT).   The only catch is that the
collisionless-sector equations must be solved with the equations
for the rest of the system iteratively.  Still, as we show, this
iteration converges quickly.  If the collisionless sector
dominates the computational effort, this iterative scheme may
provide a more computationally efficient route to a precise numerical solution.

Below we first derive the integral equations for the moments of
the massive-neutrino distribution functions and show how they
can be written as convolutions.   We then discuss aspects of the iterative
scheme \cite{Kamionkowski:2021njk} to solve the
collisionless-sector perturbations in tandem with the equations
for photons, dark matter, baryons, and gravitational potentials.
We present numerical results from a proof-of-concept code and
end with some concluding remarks.

\paragraph{Integral Solution.}
We start with the linearized collisionless Boltzmann equation in Fourier space and in synchronous gauge \cite{Ma:1995ey},\footnote{The hierarchy-less approach is equally applicable to the conformal Newtonian gauge.}
\begin{equation}\label{eqn:boltzmann}
     \frac{\partial \Psi}{\partial\tau}  + i k \mu \frac{q}{\epsilon} \Psi + \frac{d\ln f_0}{d\ln q}\left[\eta' - \frac{h'+6\eta'}{2} \mu^2 \right] = 0,
\end{equation}
and follow the notation in Ref.~\cite{Ma:1995ey} unless stated otherwise. Here the fractional phase-space-density perturbation $\Psi$ is related to the phase-space density via $f(\vec q, \vec k, \tau) = f_0(q, \tau)[1+\Psi(\vec q, \vec k, \tau)]$ with $\vec q$ being the neutrino momentum ($q \equiv |\vec q|$) and $f_0$ being the Fermi-Dirac distribution. Due to symmetry considerations \cite{Ma:1995ey}, $\Psi$ depends only on the momentum magnitude $q$, the Fourier wavenumber $k\equiv |\vec k|$, and the angle $\mu \equiv (\vec q/q) \cdot (\vec k/k)$. We have also introduced the synchronous-gauge metric perturbations $h(k,\tau)$ and $\eta(k,\tau)$, and use a prime to denote derivative with respect to conformal time $\tau$. We follow Ref.~\cite{Lesgourgues:2011rh}, thus a small deviation from Ref.~\cite{Ma:1995ey}, in defining the neutrino energy $\epsilon(q,\tau) \equiv [q^2 + a^2(\tau) m^2 / T_0^2]^{1/2}$, with $a(\tau)$ the scale factor, $m$ the neutrino mass, and $T_0$ the current neutrino temperature. We omit the arguments of all quantities if no confusion is caused.

We recognize Eq.~\eqref{eqn:boltzmann} as a first-order ODE of $\Psi$ in $\tau$, labeled by $\mu$, $q$, and $k$. Integrating this equation from some initial time $\tau_i$ to some final time $\tau_f$, we obtain the formal solution,
\begin{align}
     \Psi(\tau_f) & = e^{- i \mu k \chi(\tau_i, \tau_f)} \Psi(\tau_i) \nonumber\\ & \quad + \int_{\tau_i}^{\tau_f} e^{- i \mu k \chi(\tau, \tau_f)}  \left[- \eta' + \frac{h'+6\eta'}{2} \mu^2\right]  \frac{d\ln f_0}{d\ln q} \, d\tau.
\end{align}
Here we define the \emph{neutrino} comoving horizon $\chi(\tau_1,\tau_2; q) = \int_{\tau_i}^{\tau_f} (q/\epsilon)\, d\tau$, and omit the $q$ dependence for simpler notation. We now define the multipole moments $\Psi_l \equiv (i^l/2) \int_{-1}^{+1} \Psi(\mu) P_l(\mu)\, d\mu$ with $P_l(\mu)$ the Legendre polynomials, and use the integral representation \begin{equation}\label{eqn:spherical-bessel-integral}
     \frac{d^n}{dx^n} j_l(x) = \frac{i^l}{2} \int_{-1}^{+1} e^{-i\mu x} (-i\mu)^n P_l(\mu)\, d\mu
\end{equation}
of the spherical Bessel functions $j_l(x)$ (and its derivatives)
to arrive at the central result,
\begin{align}\label{eqn:integral-solution}
     \Psi_l(\tau_f)& = \sum_{l'=0}^\infty (-1)^{l'}(2l'+1) W_{ll'}[k \chi(\tau_i, \tau_f)]\Psi_{l'}(\tau_i) \nonumber \\
     &+ \int_{\tau_i}^{\tau_f} \frac{d\ln f_0}{d\ln q} \, d\tau \nonumber \\
 & \ \ \times \left\{- j_l[k \chi(\tau, \tau_f)] \eta' - j''_l[k \chi(\tau, \tau_f)] \frac{h'+6\eta'}{2} \right\}. \end{align}
Here, we have defined the auxiliary function,
\begin{equation}
\begin{split}
    W_{ll'}(x)  & \equiv \frac{i^{l+l'}}{2}\int_{-1}^{+1}e^{-i\mu x} P_l(\mu) P_{l'}(\mu) d\mu  \\ 
    & = i^{l'} P_{l'}\left(i\frac{d}{dx}\right)j_l(x).
\end{split}
\end{equation}

\begin{figure}
    \centering
    \includegraphics[width = \linewidth]{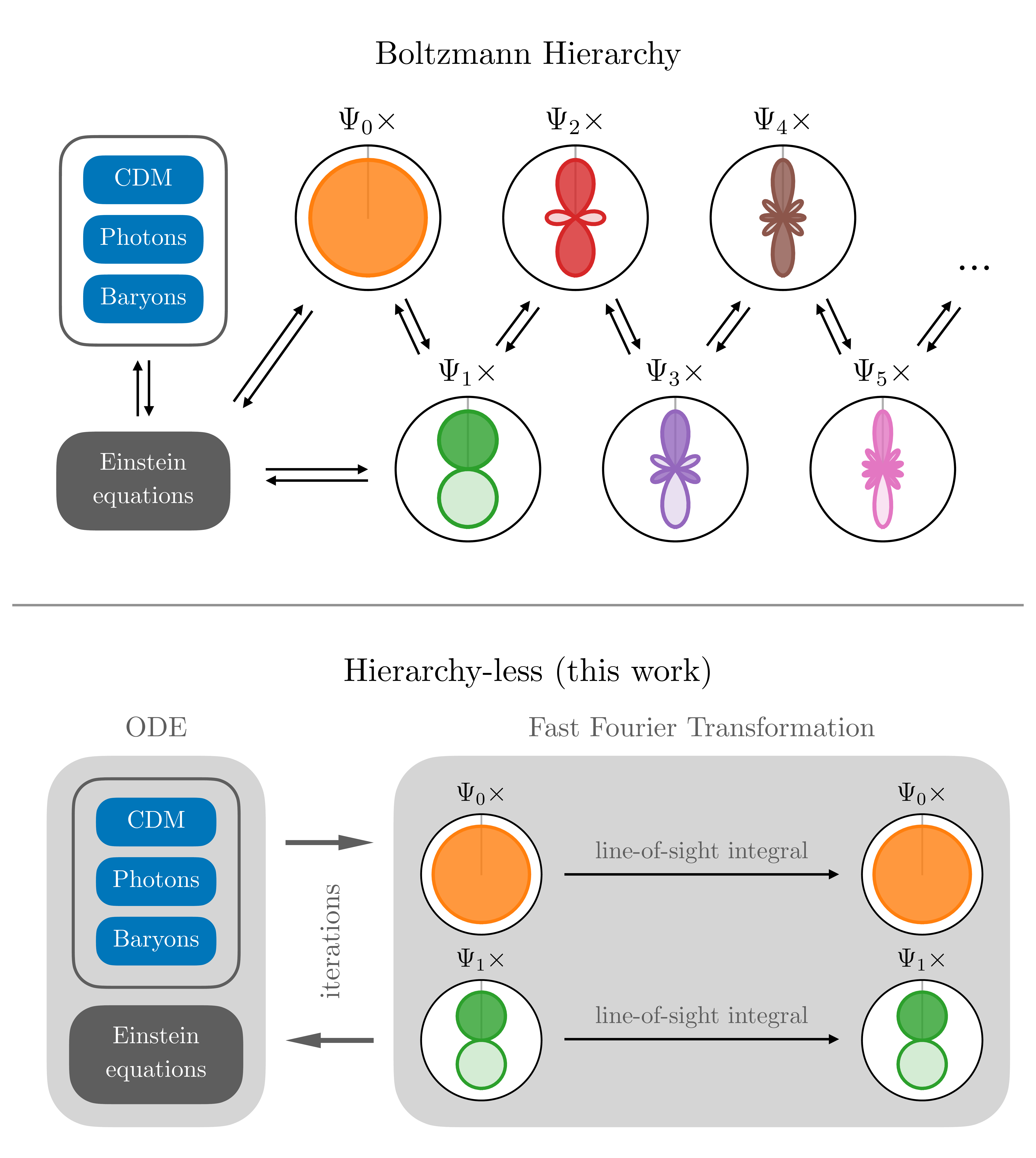}
    \caption{Comparison between the traditional solver using a truncated Boltzmann hierarchy and the new solver proposed in this paper. In the new approach, the infinite hierarchy is being replaced by two line-of-sight integrals --- for the monopole and dipole of the distribution-function --- that are evolved simultaneously with the differential equations for the rest of the system via an iterative scheme. The line-of-sight integrals are computed very efficiently via fast Fourier transforms.}
    \label{fig:bh-vs-ie}
\end{figure}

Now, we discuss the evaluation of the integral solution, Eq.~\eqref{eqn:integral-solution}.
We choose the initial time $\tau_i$ sufficiently early, ideally close to neutrino decoupling, when the higher multipoles $\Psi_l(\tau_i)$ for $l > 2$ are effectively zero. This reduces the infinite sum in Eq.~\eqref{eqn:integral-solution} to only 3 terms (i.e.\ $l'=0,1,2$). Then, $\Psi_l(\tau_f)$ for arbitrary $\tau_f > \tau_i$ can be computed by performing the integral in Eq.~\eqref{eqn:integral-solution}. Although this can be done for arbitrary $l$ too, we only need the monopole and dipole (i.e. $l=0, 1$) as those are all that appear in the Einstein equations.   A schematic comparison between the Boltzmann-hierarchy solver and the new hierarchy-less solver presented in this work is shown in Fig.~\ref{fig:bh-vs-ie}.

Although similar to the analogous integral equation for photons in Ref.~\cite{Kamionkowski:2021njk}, Eq.~\eqref{eqn:integral-solution} is different in a very important way.
The phase-space perturbation $\Psi_l(\tau)$ does \emph{not} appear inside the integral in Eq.~\eqref{eqn:integral-solution}, so Eq.~\eqref{eqn:integral-solution} is merely an integral, not a {\it bona fide} integral equation, a consequence of the fact that neutrinos are collisionless.  As we will see shortly, this allows for considerable simplification and acceleration.

\paragraph{Iterative Method}
The integrals in Eq.~\eqref{eqn:integral-solution} require the metric perturbations $h(\tau)$ and $\eta(\tau)$, but the Einstein equations that determine these quantities take as input the neutrino perturbations (as well as those of any other species).
To solve this chicken-and-egg problem, we solve the coupled system of equations iteratively, as follows.

We first choose an \emph{ansatz} for the neutrino sector, and solve the non-neutrino sector using a traditional ODE solver; then the metric perturbations are used to evaluate and update the neutrino sector via Eq.~\eqref{eqn:integral-solution}. This process is continued until some target precision is achieved. Better choice of the \emph{ansatz} enables faster convergence of the iterations. Here, we discuss several possibilities.

One simple possibility is to start with a solution to the ODEs truncating the neutrino hierarchies at
a low multipole.  These trial solutions typically take far shorter to compute compared to the full hierarchy, but nonetheless provide enough crude features in the solution for the iterative process to refine on.  The numerical results shown below are obtained with this \emph{ansatz}.

Another possibility is to use the neutrino-sector solution from the previous MCMC step as the \emph{ansatz}. A converging MCMC typically only samples fairly concentrated points around the best-fit model in the parameter space. Thus, presumably, a solution from the previous step is a very good approximation to the true solution of the current step. Along this line of reasoning, one can even maintain a small cache of certain previous MCMC steps that more or less uniformly cover the parameter space of interest. Then, in the current step, only retrieve the closest candidate as the \emph{ansatz} (although the required interpolation may be costly).  A related possibility  is to do something similar using the solutions for $\Psi_l$ from a previous $k$ value in the calculation, rescaling the conformal time so that $k\tau$ is fixed.

\paragraph{FFT Acceleration}

The line-of-sight integral can be written as a convolution between a cosmology-independent kernel and the metric perturbations.  The integral in Eq.~\eqref{eqn:integral-solution} can be written schematically as
\begin{equation}
    I(\tau_f) = \int _{\tau_i} ^{\tau_f} F(\tau) K[x(\tau_f) - x(\tau)] \, d\tau.
\end{equation}
For the first term in the integral in Eq.~\eqref{eqn:integral-solution},
\begin{equation}
    F(\tau) \equiv -\eta' \frac{d\ln f_0}{d\ln q} \quad {\rm and} \quad K(x) = j_l(x), 
\end{equation}
and for the second term in the integral in Eq.~\eqref{eqn:integral-solution},
\begin{equation}
    F(\tau) \equiv -\frac{h'+6\eta'}{2} \frac{d\ln f_0}{d\ln q} \quad {\rm and} \quad K(x) = j_l''(x),
\end{equation}
but the following derivation applies to both cases. We define $x(\tau) \equiv k\chi(\tau_i, \tau)$ and we have used the fact that these distances are additive; i.e.\ $\chi(\tau_i, \tau) + \chi(\tau, \tau_f) = \chi(\tau_i, \tau_f)$. Now, we change the integration variable using the inverse function $\tau = \tau(x)$ and $d\tau/dx = \epsilon(x)/(qk)$, giving
\begin{equation}
    I[\tau(x_f)] = \int _0 ^{x_f} \frac{\epsilon(x)}{qk} F[\tau(x)] K(x_f - x) \, dx,
\end{equation}
where $x_f \equiv k\chi(\tau_i,\tau_f)$. Defining the function $G(x) \equiv \epsilon(x) F[\tau(x)] / (qk)$, we have
\begin{equation}
    I[\tau(x_f)] = \int _0 ^{x_f} G(x) K(x_f - x) \, dx = (G \star K)(x_f).
\end{equation}
Here $G \star K$ denotes the \emph{Laplace} convolution between $G$ and $K$. The discrete samples of $I[\tau(x_f)]$ can be computed from the discrete samples of $G(x)$ and $K(x)$ very efficiently via FFT. Note that the $x$-samples (or $\tau$-samples) do not need to be uniform, in which case the non-uniform FFT can be used without impacting the $O(N\log N)$ complexity.

\begin{figure*}
    \centering
    \includegraphics[width=\linewidth]{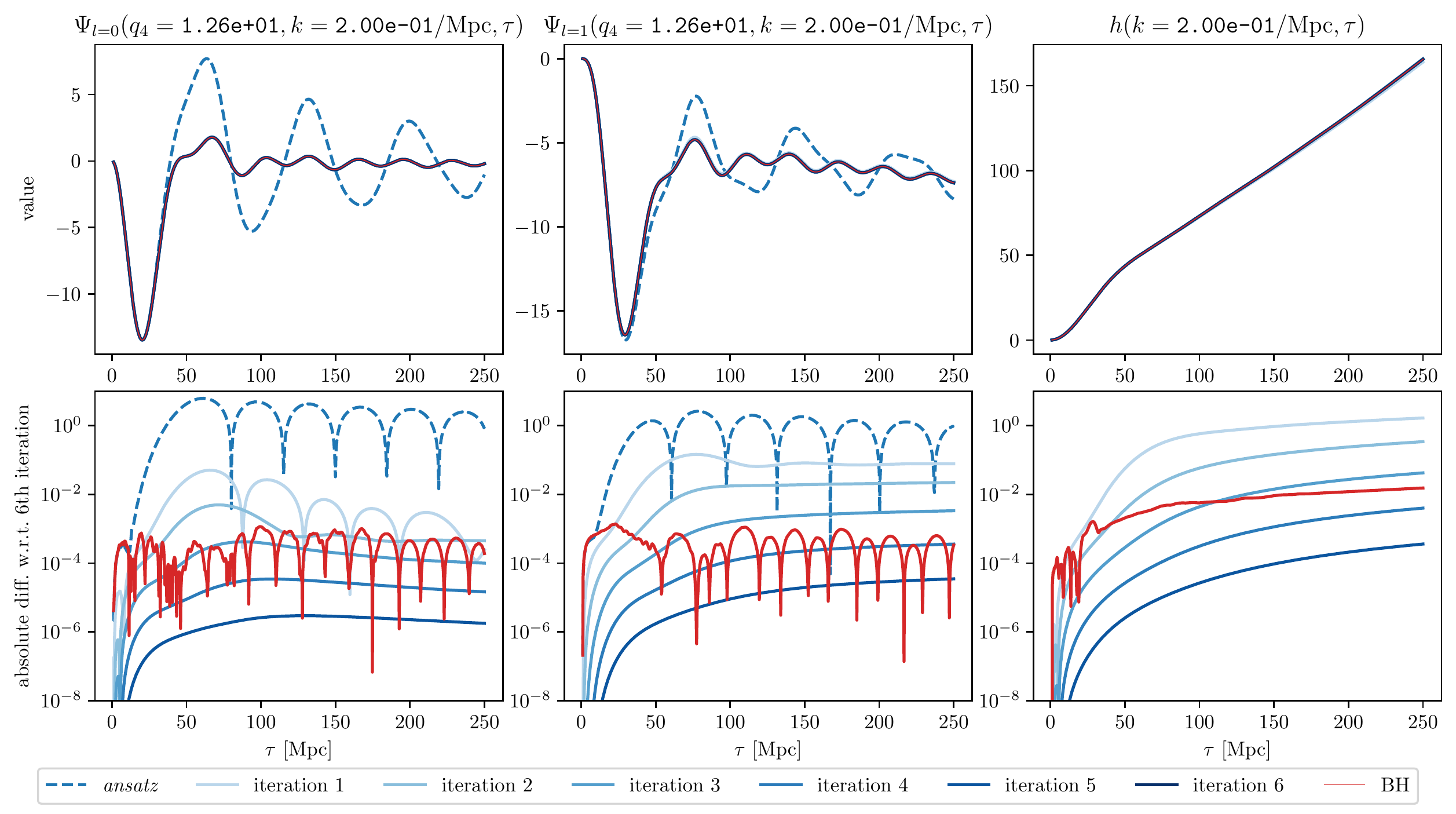}
    \caption{Numerical results of the hierarchy-less solver. We show (from left to right) the evolution of the neutrino distribution-function monopole $\Psi_0$, dipole $\Psi_1$, and the synchronous-gauge metric perturbation $h$. The top panels show the initial \emph{ansatz} (blue dashed, obtained by solving a very short hierarchy cut at $l=3$), the results of 6 iterations (light blue to dark blue, solid), and the solution obtained from the full Boltzmann hierarchy (red solid). Note that the iterations are overlapping due to the rapid convergence. Each bottom panel shows the absolute differences between the lines in the corresponding top panel comparing to the results of the $6^{\rm th}$ iteration.}
    \label{fig:numerical}
\end{figure*}

\begin{figure}
    \centering
    \includegraphics[width=\linewidth]{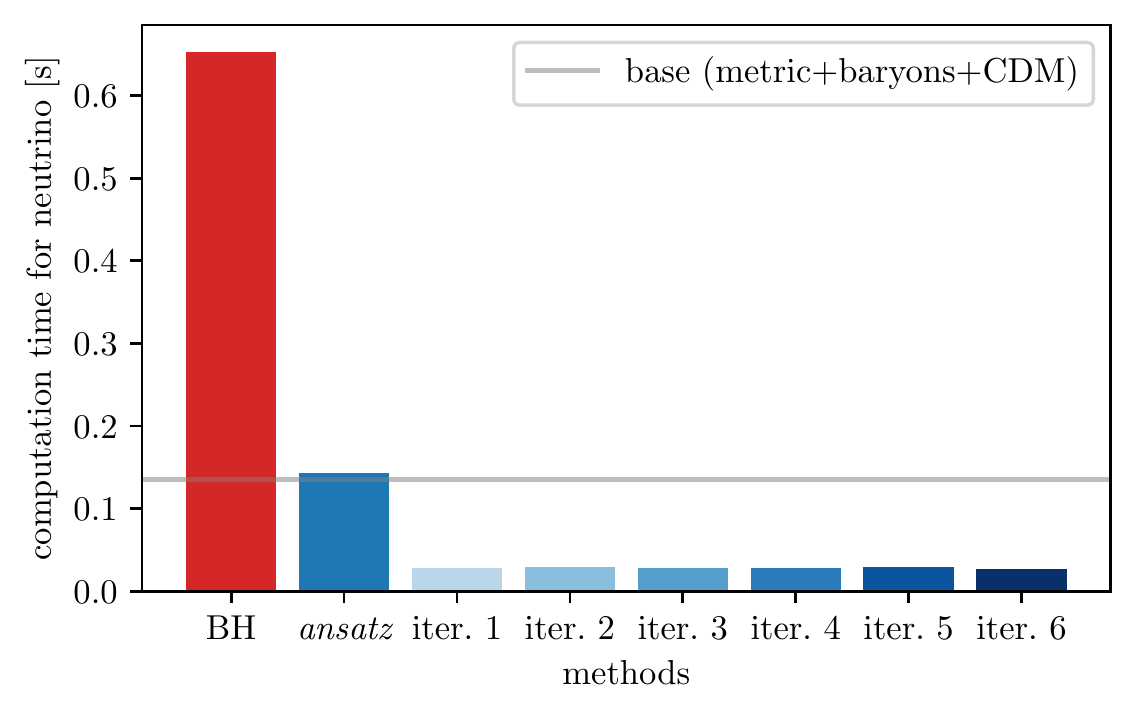}
    \caption{Comparison of computation time \emph{for neutrinos} in obtaining Fig.~\ref{fig:numerical}. Here, the computation time is defined to be the total time spent on the neutrino hierarchy (for the Boltzmann-hierarchy case and for obtaining the \emph{ansatz}) or the neutrino line-of-sight integral (for the iterations). As a reference, we plot the computation time for baryons, CDM, and metric as a gray horizontal line.}
    \label{fig:benchmark}
\end{figure}

\paragraph{Numerical Demonstrations}
Our calculation proceeds as follows.  (1) We first solve the complete set of ODEs for the baryons, dark matter, photon moments, gravitational potentials, and neutrinos.  However, we truncate all the {\it neutrino} Boltzmann hierarchies at $l=3$.  This then provides an initial solution for the potentials $h(\tau)$ and $\eta(\tau)$.  (2) We then evaluate the neutrino monopoles $\Psi_{0}(\tau)$ and dipoles $\Psi_{1}(\tau)$ for all momenta $q$ from Eq.~\eqref{eqn:integral-solution}, using the FFT method described above.  (3) We then go back and solve the ODEs for the baryons, dark matter, photon moments, and gravitational potentials.  However, this time we use the results of step (2) for the neutrino source terms in the Einstein equations.  (4) We then iterate steps (2) and (3) until the desired precision in the neutrino moments or the gravitational potentials are achieved.

For step (2), one could alternatively simply evaluate the integral equation for either the monopole {\it or} the dipole (rather than both) and then obtain the other from the continuity equation.  We have found, though, that the solutions converge more rapidly if they are both evaluated with the integral equation, with little additional computational effort.

We develop a proof-of-concept \texttt{python} code to demonstrate the potential of the new hierarchy-less solver. We adopt a $\Lambda$CDM cosmology with one species of massive neutrino with $m_\nu = 0.06\, {\rm eV}$.
The cosmological parameters are chosen to be the default in \texttt{CLASS v3.0.1}. As an example, we solve the $k=0.2\, {\rm Mpc}^{-1}$ mode in the conformal-time interval $\tau\in [1, 250]\, {\rm Mpc}$, and discretize the $q$-integration with 5 Gauss-Laguerre nodes. We choose $\tau_{\rm max}=250$ Mpc so that $k\tau_{\rm max} = 50$, which is significantly larger than the standard values to switch on the fluid approximation for the non-cold collisionless relics (e.g., the standard value  in \texttt{CLASS} is 31).

In Fig.~\ref{fig:numerical}, we demonstrate the rapid convergence of the iterative process, and the accuracy of the converged solution. Here, we construct the \emph{ansatz} by solving the system with a short neutrino hierarchy truncated at $l=3$, and iterate 6 times from that. We then compare the result from the last iteration with the Boltzmann-hierarchy approach truncated at $l=30$. In each iteration, we compute the neutrino line-of-sight integral via an FFT of $N=1024$ points. Whenever there is a need to solve ODEs, we use the \texttt{RK45} adaptive integrator with $\texttt{rtol}=10^{-4}$ and $\texttt{atol}=10^{-8}$.

In Fig.~\ref{fig:benchmark}, we compare the computation time \emph{for neutrinos} in obtaining Fig.~\ref{fig:numerical}, defined to be the total time spent on the neutrino hierarchy (for the Boltzmann-hierarchy case and for obtaining the \emph{ansatz}) or on the neutrino line-of-sight integral (for the iterations). The time for the \emph{ansatz} can be eliminated if we obtain the \emph{ansatz} form the previous MCMC step, or a previous $k$. The time for each iteration is expected to scale as $O(N\log N)$.

\paragraph{Conclusions}
We have shown that each of the Boltzmann hierarchies for collisionless species can be replaced by a set of integrals that can be evaluated efficiently with FFT, but at the price of solving the equations for the rest of the system iteratively.  Even so, our simple numerical experiments suggest that the iteration can converge quickly with even a simple initial {\it ansatz} and thus hold the prospect to accelerate cosmological-perturbation calculations, especially in models with multiple mass eigenstates.

Moreover, we emphasize that the new approach described in this work can be used to accelerate models with other non-cold collisionless species~\cite{Deramo:2020gpr,Das:2021pof,Kunze:2021qxt,Decant:2021mhj,Abellan:2020pmw}, without much adaptation. It should also apply to scenarios where these (or the neutrino) species have non-thermal homogeneous distribution function $f_0$ \cite{Alvey:2021sji}. In general, we expect the acceleration to be more significant with a larger non-cold collisionless sector.  Still, the optimization of the computational efficiency subject to some precision threshold is a difficult problem, both for the traditional approach and the one we are suggesting here.  It will require more work to determine more conclusively whether this can be implemented to improve the performance while providing the type of reliability and flexibility available with current codes.

\smallskip

We thank V.\ Poulin for useful discussions.  This work was supported by the Simons Foundation and by National Science Foundation grant No. 2112699. JLB was supported by the Allan C. and Dorothy H. Davis Fellowship.


\begin{thebibliography}{99}
\bibitem{Lewis:1999bs}
A.~Lewis, A.~Challinor and A.~Lasenby,
``Efficient computation of CMB anisotropies in closed FRW models,''
Astrophys. J. \textbf{538}, 473-476 (2000)
[arXiv:astro-ph/9911177 [astro-ph]].

\bibitem{Lesgourgues:2011rh}
J.~Lesgourgues and T.~Tram,
``The Cosmic Linear Anisotropy Solving System (CLASS) IV: efficient implementation of non-cold relics,''
JCAP \textbf{09}, 032 (2011)
[arXiv:1104.2935 [astro-ph.CO]].

\bibitem{earlywork}
   R.~A.~Sunyaev and Y.~.B.~Zeldovich,
  ``Small scale fluctuations of relic radiation,''
  Astrophys.\ Space Sci.\  {\bf 7}, 3 (1970);
  P.~J.~E.~Peebles and J.~T.~Yu,
  ``Primeval adiabatic perturbation in an expanding universe,''
  Astrophys.\ J.\  {\bf 162}, 815 (1970);
  J.~Silk,
  ``Fluctuations in the Primordial Fireball,''
  Nature \textbf{215}, no.5106, 1155-1156 (1967);
  J.~R.~Bond and G.~Efstathiou,
  ``Cosmic background radiation anisotropies in universes dominated by nonbaryonic dark matter,''
  Astrophys.\ J.\  {\bf 285}, L45 (1984);
  J.~R.~Bond and G.~Efstathiou,
  ``The statistics of cosmic background radiation fluctuations,''
  Mon.\ Not.\ Roy.\ Astron.\ Soc.\  {\bf 226}, 655 (1987);
  M.~L.~Wilson and J.~Silk,
  ``On the Anisotropy of the cosmological background matter and radiation distribution. 1. The Radiation anisotropy in a spatially flat universe,''
  Astrophys.\ J.\  {\bf 243}, 14 (1981);
  N.~Vittorio and J.~Silk,
  ``Fine-scale anisotropy of the cosmic microwave background in a universe dominated by cold dark matter,''
  Astrophys.\ J.\  {\bf 285}, L39 (1984);
  U.~Seljak and M.~Zaldarriaga,
  ``A Line of sight integration approach to cosmic microwave background anisotropies,''
  Astrophys. J. \textbf{469}, 437-444 (1996);
  [arXiv:astro-ph/9603033 [astro-ph]];
F.~Y.~Cyr-Racine and K.~Sigurdson,
``Photons and Baryons before Atoms: Improving the Tight-Coupling Approximation,''
Phys. Rev. D \textbf{83}, 103521 (2011)
[arXiv:1012.0569 [astro-ph.CO]].

\bibitem{Green:2019glg}
D.~Green, M.~A.~Amin, J.~Meyers, B.~Wallisch, K.~N.~Abazajian, M.~Abidi, P.~Adshead, Z.~Ahmed, B.~Ansarinejad and R.~Armstrong, \textit{et al.}
``Messengers from the Early Universe: Cosmic Neutrinos and Other Light Relics,''
Bull. Am. Astron. Soc. \textbf{51}, no.7, 159 (2019)
[arXiv:1903.04763 [astro-ph.CO]].

\bibitem{Deramo:2020gpr}
F.~D'Eramo and A.~Lenoci,
``Lower mass bounds on FIMP dark matter produced via freeze-in,''
JCAP \textbf{10}, 045 (2021)
[arXiv:2012.01446 [hep-ph]].

\bibitem{Das:2021pof}
S.~Das, A.~Maharana, V.~Poulin and R.~Kumar,
``Non-thermal hot dark matter in light of the $S_8$ tension,''
[arXiv:2104.03329 [astro-ph.CO]].

\bibitem{Kunze:2021qxt}
K.~E.~Kunze,
``CMB anisotropies and linear matter power spectrum in models with non-thermal neutrinos and primordial magnetic fields,''
JCAP \textbf{11}, no.11, 044 (2021)
[arXiv:2106.00648 [astro-ph.CO]].

\bibitem{Decant:2021mhj}
Q.~Decant, J.~Heisig, D.~C.~Hooper and L.~Lopez-Honorez,
``Lyman-$\alpha$ constraints on freeze-in and superWIMPs,''
[arXiv:2111.09321 [astro-ph.CO]].

\bibitem{Abellan:2020pmw}
G.~F.~Abellan, R.~Murgia, V.~Poulin and J.~Lavalle,
``Hints for decaying dark matter from $S_8$ measurements,''
[arXiv:2008.09615 [astro-ph.CO]].

\bibitem{Alvey:2021sji}
J.~Alvey, M.~Escudero and N.~Sabti,
``What can CMB observations tell us about the neutrino distribution function?,''
[arXiv:2111.12726 [astro-ph.CO]].


\bibitem{integralequations}
  S.~Weinberg,
  ``Damping of tensor modes in cosmology,''
  Phys. Rev. D \textbf{69}, 023503 (2004)
  [arXiv:astro-ph/0306304 [astro-ph]];
  D.~Baskaran, L.~P.~Grishchuk and A.~G.~Polnarev,
  ``Imprints of Relic Gravitational Waves in Cosmic Microwave Background Radiation,''
  Phys. Rev. D \textbf{74}, 083008 (2006)
[arXiv:gr-qc/0605100 [gr-qc]];
  R.~Flauger and S.~Weinberg,
  ``Tensor Microwave Background Fluctuations for Large Multipole Order,''
  Phys. Rev. D \textbf{75}, 123505 (2007)
  [arXiv:astro-ph/0703179 [astro-ph]];
  J.~R.~Pritchard and M.~Kamionkowski,
  ``Cosmic microwave background fluctuations from gravitational waves: An Analytic approach,''
  Annals Phys. \textbf{318}, 2-36 (2005)
  [arXiv:astro-ph/0412581 [astro-ph]];
  S.~Weinberg,
  ``A No-Truncation Approach to Cosmic Microwave Background Anisotropies,''
  Phys. Rev. D \textbf{74}, 063517 (2006)
  [arXiv:astro-ph/0607076 [astro-ph]].

\bibitem{Kamionkowski:2021njk}
M.~Kamionkowski,
``Cosmological perturbations without the Boltzmann hierarchy,''
Phys. Rev. D \textbf{104}, no.6, 063512 (2021)
[arXiv:2105.02887 [astro-ph.CO]].

\bibitem{Ma:1995ey}
C.~P.~Ma and E.~Bertschinger,
``Cosmological perturbation theory in the synchronous and conformal Newtonian gauges,''
Astrophys. J. \textbf{455}, 7-25 (1995)
[arXiv:astro-ph/9506072 [astro-ph]].

\bibitem{Blas:2011rf}
D.~Blas, J.~Lesgourgues and T.~Tram,
``The Cosmic Linear Anisotropy Solving System (CLASS) II: Approximation schemes,''
JCAP \textbf{07}, 034 (2011)
[arXiv:1104.2933 [astro-ph.CO]].





\end{thebibliography}
\end{document}